\documentstyle[aps,prl,epsf,multicol]{revtex}

\begin{document}
\title{Hall Voltage Fluctuations as a Diagnostic of
Internal Magnetic Field Fluctuations 
in  High Temperature Superconductors and the Half-filled 
Landau Level}
\author{L. B. Ioffe}
\address{Physics Department, Rutgers University, Piscataway, NJ 08855 \\
and Landau Institute for Theoretical Physics, Moscow}
\author{G. B. Lesovik}
\address{Institute for Solid State Physics, RAS, Chernogolovka, 142432, Russia}
\author{A. J. Millis}
\address{Bell Laboratories, Lucent Technologies, Murray Hill, NJ 07974}
\maketitle
\begin{abstract}
Fluctuations of the Hall voltage reveal information about long wavelength
magnetic field fluctuations. 
If gauge theories of strongly correlated electrons are correct,
such fluctuations are particularly large in the half-filled
Landau level and in high $T_c$ superconductors.
We present estimates for the magnitude, system size and frequency dependence
of these fluctuations. 
The frequency dependence contains information about instantons in the gauge
field. 
\end{abstract}

\begin{multicols}{2}

It is widely believed that the anomalous properties of high $T_c$
superconductors imply that a soft mode exists, but there is as yet no
consensus on the origin of this mode.  
One possibility is an internal gauge field associated with spin-charge 
separation \cite{Baskaran87,Ioffe89}.
In this communication we show that measurements of the fluctuations in the
Hall voltage at zero applied magnetic field and non-zero applied current can 
reveal the presence of internal gauge fluctuations. 
The idea is simply that the internal gauge fluctuations produce an effective
magnetic field which affects the motion of electrons in the same way as a
conventional magnetic field.
In the presence of a current, fluctuations in the effective field 
will therefore
lead to fluctuations in the voltage transverse to the current.
Experimental detection of the fluctuations would confirm the validity of the
gauge field description while their frequency dependence contains information 
about dynamics of the gauge field which is difficult to obtain by other means.

The internal gauge field is associated with the spin-charge separation; within
the gauge theory approach it occurs at $T>T_c$ for optimally doped and
underdoped materials.
For overdoped materials spin-charge separation may exist at high $T$, but
there is a crossover to Fermi liquid like behavior with smaller gauge field
fluctuations at $T_{FL}>T_c$ \cite{Lee90}
As we shall discuss below, the fluctuations should be most easily observable
in moderately underdoped materials at temperatures near $T_c$, but
sufficiently far above it that conventional superconducting fluctuations are
not observable.

If spin-orbit coupling is important, long wavelength fluctuations of the
electron spin degrees of freedom will lead to similar effects which we  
show are much smaller in magnitude.

Rather similar considerations apply to the half-filled Landau level, where a
Chern-Simons gauge theory has been proposed  to describe the low energy
physics\cite{Kalmeyer92,Halperin93} but one must measure fluctuations
relative to the background Hall voltage caused by the magnetic field which
created the $\nu = 1/2$ state. 
Also the dynamics of the gauge field is somewhat different as discussed below.

We suppose that there exists a dimensionless field $\phi$ which has
significant long wavelength fluctuations and which induces an internal field
\begin{equation}
B_{i nt} (\vec{r},t) \equiv f \phi (\vec{r},t)
\label{B_int}
\end{equation}
which we assume affects the motion of electrons in the same way as does a
magnetic field. 
Because we are interested in long wavelength fluctuations we take the relation
between $\phi$ and B to be local and specified by a constant, $f$, which has
the dimension of magnetic field. 

In the U(1) gauge theory of high $T_c$ materials or of the half-filled Landau
level, $\phi$ is essentially the internal magnetic field $h = \nabla  \times
{\bf a}$ induced by the gauge field $\bf a$. 
Because $h$ has dimension $[length]^{-2}$ we make it dimensionless by
multiplying by two factors of a microscopic length $l_0$.  
Different choices of $l_0$ correspond to different values of the coupling
constant $f$. 
For high $T_c$ materials it is natural to take $l_0$ to be the lattice
constant, $4 \AA$; for the half-filled Landau level  the natural choice is 
the interparticle spacing, which is of order $300 \AA$.
In both cases $f= \hbar c/(el_0^2)$.  
For high $T_c$ materials
($l_0 =4 \AA$) $f \sim 4 \times 10^7 gauss$; for the half-filled
Landau level ($l_0 = 300 \AA$) $f \sim 8 \times 10^3 gauss$.

The electron spin degrees of freedom are an additional source of magnetic
field fluctuations.  
In this case $\phi$ is the spin polarization per unit cell  and $f = \mu_B /
\lambda^3$ where $\lambda$ is a length. 
If the spin polarization were spread uniformly over the unit cell one would
take $\lambda \sim l_0$, leading to $f \sim 125$ gauss for high $T_c$
materials.  
However, in high $T_c$ materials the conduction-band states involve Cu
$d_{x^2-y^2}$ orbitals and the associated large spin-orbit interaction
will increase the coupling of the spin fluctuations to the orbital motion of
the electrons by a factor which we estimate as follows.  
The Cu d-orbitals are of size $r_{Cu} \sim 0.5 \AA \sim l_0/10$.   
Because the only d-orbital with appreciable weight at the fermi surface is the
$d_{x^2-y^2}$ orbital, the effect on the motion of the electrons is
controlled by the flux per unit cell, which for a dipole of size $r_{Cu}$  is
$\mu_B/r_{Cu}$ implying an enhancement of $f$ by one factor of
$l_0/r_{Cu} \sim 10$. 
Thus, for spin effects in high $T_c$ materials $f \sim 10^3
gauss$, four orders of  magnitude smaller than the $f$ for gauge fields.

\vspace{-.75in}
\centerline{\epsfxsize=8cm \epsfbox{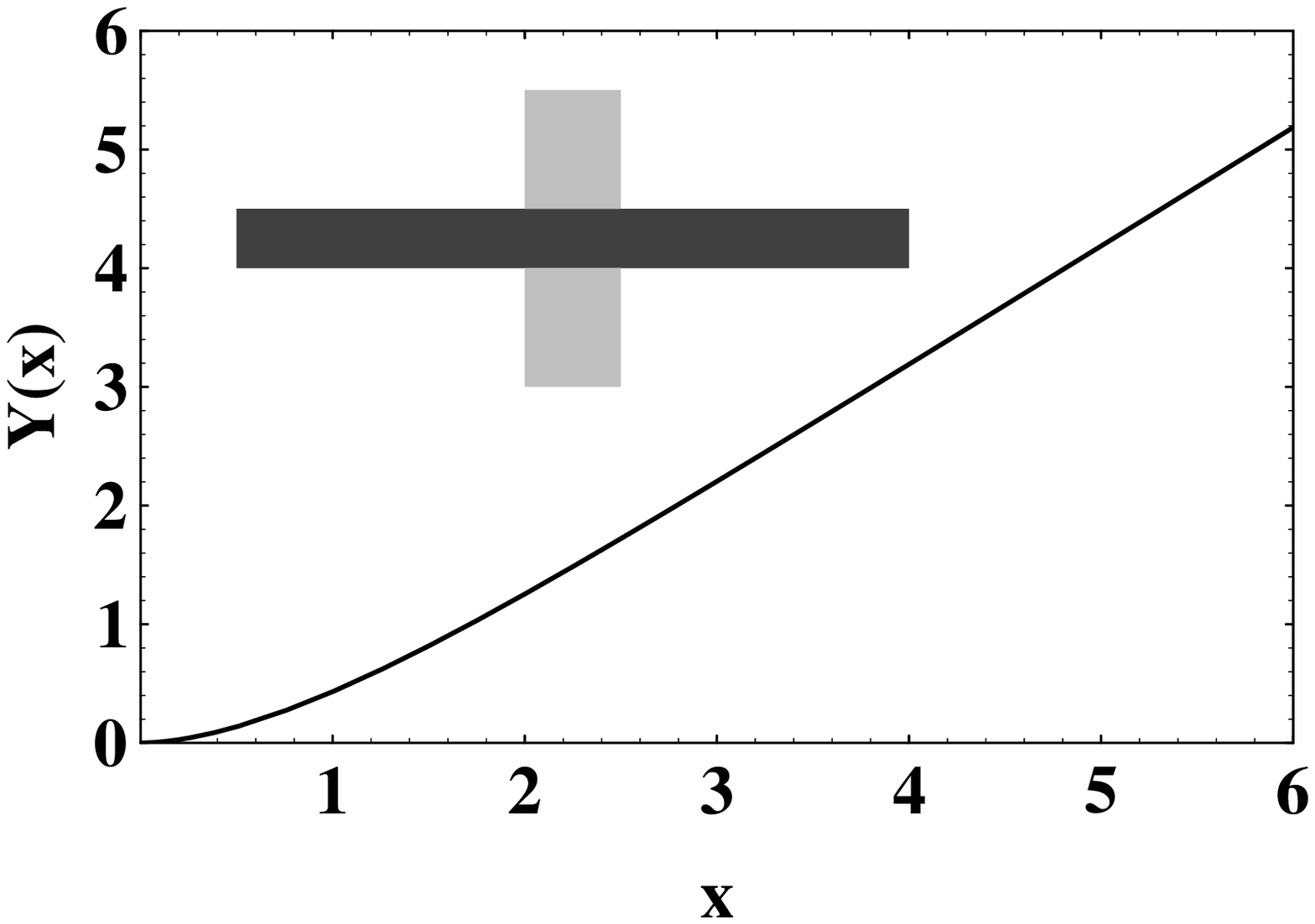}}
\vspace{-.75in}
{ \small
Fig 1.
The dimensionless intensity $Y(x)$ of the noise as a function of $x=2L_x/L_y$
(see Eq. (\ref{V^2})).  Inset:    schematic of proposed experiment.
The sample is the heavily shaded bar, the current, $I$, flows
along the long axis (x-direction).  The contact pads are
the lighly shaded rectangles; they have
width  $L_x$ and the transverse distance from one to the
other is $L_y$.
}
\vspace{.1in}

We now consider the dynamics of $\phi$.
We are interested in long wavelengths and low frequencies; we therefore assume
the dynamics is diffusive, with diffusion constant $D$, but we allow for
non-conservation of $\phi$ which we parametrize by a relaxation rate $\gamma$
due physically to instantons in the gauge field case \cite{Polyakov87} and
spin-orbit coupling in the spin case.  
We assume that the relevant physics is two dimensional.  
Thus, we take the infinite-system form of the $\phi - \phi$ correlator to be:
\begin{equation}
D^{(R)}(q,\omega) \equiv \langle [ \phi , \phi ] \rangle_{q,\omega} =
 \Lambda \; 
\frac{Dq^2 + \gamma}{-i\omega + Dq^2 + \gamma} .
\end{equation}
Here $\Lambda$ is essentially the susceptibility  of the field
conjugate to $\phi$; it has the dimension of $(length)^2/energy$.  

The thermal and quantum fluctuations of $\phi$ are given as usual by 
$\langle \phi \phi \rangle_{q,\omega} = coth \frac{\omega}{2T} Im D(q,\omega)$.
Knowing the fluctuations of $\phi$ we may calculate the fluctuations of $B$ and
thus the fluctuations of the Hall voltage. 
For definiteness we discuss the situation shown in the inset to Fig. 1.
The fundamental equation is
\begin{equation}
\vec{\jmath} = \mbox{\boldmath $\sigma$} \vec{E}
\label{j}
\end{equation}
where the current density $\vec{\jmath}$ is the sum of the imposed current
density $\jmath_x=I/L_y$ and a transverse part $\vec{\jmath}_T$ whose $y$
component vanishes at the upper and lower edges of the strip; the electric
field $\vec{E}$ is  longitudinal and because of the fluctuating magnetic field
$\mbox{\boldmath $\sigma$}$ is not diagonal.  
Because $B_{int}$ is always small, we use the linear approximation
\begin{equation}
\mbox{\boldmath $\sigma$}^{-1} = \rho_{xx} {\bf 1} -
\mbox{\boldmath $\epsilon$} \rho_H B_{i nt}
\label{sigma}
\end{equation}
Here $\mbox{\boldmath $\epsilon$}$ is the antisymmetric tensor, $\rho_{xx}$ is
the usual dc resistivity, and $\rho_H$ the usual Hall resistivity.

Equations (\ref{j}) and (\ref{sigma}) 
may be solved to determine the potential difference $V_H(x)$
between the upper and lower edges of the strip at position $x$.
It is more convenient to write a formula for
$V^H_q = \int dx \; e^{iqx} V_H(x)$:
\begin{equation}
V^H_q = \frac{\rho_HI}{L_y} \int_{-\frac{L_y}{2}}^{\frac{L_y}{2}} \!\! dy dy'
[\delta (y-y^{\prime}) - q^2 K(y,y^{\prime})] B_{int}(q,y)
\label{V_H(q)}
\end{equation}
The kernel $K$ expresses the effect of the boundary
condition $\vec{\jmath} (y= \pm L_y/2) =0$ and may be written
\begin{equation}
\begin{array}{ll}
2q & K(y,y^{\prime}) = exp - q|y-y^{\prime} | \\
   & - exp [-qL_y/2] \;[
\frac{sinh \; qy \;  sinh \; qy^{\prime}}{sinh \; qL_y/2}
+ \frac{cosh \; qy \;cosh \; qy}{cosh \; qL_y/2}]
\end{array}
\end{equation}
The important values of $q$ are $q \sim 1/L_x$.  
$K(y,y')$ suppresses the contributions of wavevectors $q > 1/L_y$; if
the experimental system has $L_x \gtrsim L_y$ $K$ may be
neglected in Eq. (\ref{V_H(q)}).

These arguments must be slightly modified for the gauge field case.
Here one has "spinons" (s) and "holons" (h) coupled by the internal gauge
field. 
The hydrodynamic equations are \cite{Ioffe89,Ioffe90}
\begin{eqnarray}
\vec{\jmath}_s &= \mbox{\boldmath $\sigma$}_s \vec{e} \\
\vec{\jmath}_h &= \mbox{\boldmath $\sigma$}_h (\vec{e}+\vec{E})
\end{eqnarray}
The internal electric field $e$ and the physical electric field $E$
are longitudinal, and fixed by the requirements
$\vec{\jmath}_h = I/L_x + \vec{\jmath}_{h,T}$,
$\vec{\jmath}_s = -I /L_x+ \vec{\jmath}_{s,T}$,
and $\jmath^y_{h,s} (y=\pm L/2)=0$.
Solving these equations leads to Eq. (\ref{V_H(q)}) with $\rho_H$ replaced
by $\rho_H^s + \rho_H^h$.
As discussed in Ref. \cite{Ioffe90}, the quantity $\rho_H^s + \rho_H^h$ this
is not the physical Hall resistivity, but is  of the same order of magnitude. 

Further, in the gauge field problem the gauge field propagator $D$ is affected
by the finite geometry.  
This may be seen from the fact that the inverse gauge-field propagator is the
sum of two terms \cite{Ioffe89}, one coming  from the diamagnetic
susceptibility energy density $\frac{1}{2}  \; \chi_{dia} h^2$, and the other
from the fact that the gauge fluctuations induce particle currents, which
dissipate. 
This latter term is affected by boundary conditions and in the hydrodynamic
limit  may be calculated as $V_H(q)$ above. 
We find the formal expression
\begin{eqnarray}
\langle \phi(q,y) \phi(q,y^{\prime}) \rangle &=&
l_0^4 \langle h(q,y) h(q,y^{\prime}) \rangle \nonumber \\
&=& coth \left ( \frac{\omega}{2T} \right ) \; \Lambda \;
\frac{(\omega (D\hat{K}^{-1} + \gamma)}
{\omega^2 + (D\hat{K}^{-1} + \gamma)^2}
\label{phi_phi}
\end{eqnarray}
Here $\hat{K}^{-1}$ is the inverse of the operator $\hat{K}$ given above,
$\Lambda \; =l_0^4/ \chi_{dia}$ and $D=\chi_{dia}/\tilde{\sigma}$,
$\tilde{\sigma}=p_0l/(2\pi)$ with $l$ the mean free path and $p_0$ the
curvature of the Fermi surface.   
Note that if $L_y \rightarrow \infty$ $\hat{K}^{-1} \rightarrow q^2$.  
If also $\gamma \rightarrow 0$ then $\langle h(q) h(q) \rangle$ tends to the
familiar form $coth \frac{\omega}{2T} \frac{q^2 \omega \tilde{\sigma}}
{(\tilde{\sigma} \omega)^2+ (\chi_{dia}q^2)^2}$.
The estimate $\chi_{dia}/l_0^2 \sim 500K$ has been obtained
from an analysis of the resistivity.  \cite{Ioffe90}
Using this value and a mean free path of $100 \AA$ at T=100K
we find $D \sim 1 \times 10^{-2}$ $cm^2/sec$.
In Eq. (\ref{phi_phi}) we have added the cutoff $\gamma$ by hand; 
the formula is correct in the two limits $\gamma \gg D/L_y^2$ and in
$\gamma \ll D/L_y^2$ but may not be quantitatively accurate
in the crossover regime $L_y^2 \sim D/\gamma$.

In the gauge field case, $\gamma$ is due to instantons;
these are tunnelling processes in which the flux due to the gauge
magnetic field changes by $2\pi$.
The action of an instanton was shown \cite{Ioffe89} to diverge as
$\omega^{-1/3}$ at $T=0$, so quantum fluctuations of instantons are
completely suppressed.
Thermally assisted tunnelling is allowed; the rate per unit area $l_0^2$  may
be roughly estimated by using $T$ to cut off the divergence found in the
quantum regime, yielding $\gamma \sim T (Tl^2/\chi_{dia})^{2/3} 
\exp\left[ -\alpha (\chi_{dia}p_0^2/\pi^2 T )^{1/3} \right]$ with 
$\alpha \sim 3.5$.
Thus one expects that $\gamma$ is small enough that it is not important for
calculations of energetics or microscopic properties such as resistivities,
but it may well be significant on the length and time scales
relevant for noise experiments. 
The dynamics of instantons is ``telegraph'' like causing a change of $2\pi$
in the flux 
over a time scale $\sim 1/T$, while the time between instantons in the 
same plaquette is $\sim \gamma^{-1} \gg 1/T$.
Each instanton changes the average internal field by $2\pi/L_xL_y$ which is
small compared to the typical value of the internal field for reasonable
sample sizes. 

The above estimates for $\chi_{dia}$ and $\gamma$ were obtained on the basis
of the uniform RVB model which is expected to describe the normal state of
optimally doped materials \cite{Lee90}
In the overdoped materials a crossover to a Fermi liquid regime occurs at
$T_{FL} > T_c$ \cite{Lee90}; in the present formalism this manifests itself as
a divergence of $\chi_{dia}$ which implies $\Lambda \rightarrow 0$.
In underdoped materials, a pseudogap is formed at a temperature $T_{PG}>T_c$.
Within the spin-liquid approach this is due to a BCS-like pairing of
spinons \cite{pseudogap} which produces an increase in $\chi_{dia}$ and
$\tilde{\sigma}$ thereby reducing $\Lambda$ and $\gamma$. 
In the underdoped materials the formation of the pseudogap changes bulk
properties such as the spin susceptibility or $d\rho/dT$ only modestly so we
expect that $\chi_{dia}$ and $\tilde{\sigma}$ and thus the total noise power do
not decrease much.
However, $\gamma$ depends exponentially on parameters, so the initial effect
of the underdoping will be to dramatically decrease the characteristic noise
frequency.

To obtain analytical formulas for the potential difference fluctuations 
we use Eq. (\ref{V_H(q)}) to write an equation for $V^2$ and then use
Eqs. (\ref{B_int}) and (\ref{phi_phi}) to average over the field fluctuations.
The resulting expressions are cumbersome in general but simplify for
relatively large samples, $L^2 \gg D/\gamma$.
In this case the local field is effectively $\delta$-correlated in space and
we get for the potential averaged over a distance  $L_x$
\begin{equation}
\langle V^2 \rangle_\omega = 2\pi \frac{T \gamma\Lambda f^2}{\omega^2+\gamma^2}
\left( \frac{\rho_H I }{L_x} \right)^2 Y\left(\frac{2L_x}{L_y}\right)
\label{V^2}
\end{equation}
where $Y(x)$ is a dimensionless function which depends on the shape of the
contact pads.
For square pads shown in Fig. 1 we find the function shown in Fig. 1 with the 
following limiting values:
$Y(x) \approx 1/(2\pi) \ln(1/x) x^2$ at $x\ll 1$, $Y(1)=0.4316$, $Y(x)=x$
at $x\gg 1$.

In the opposite limit, $L^2 \ll D/\gamma$, we make estimates.
We assume $L_x \sim L_y = L$ and neglect the details of the boundary
effects, set $K^{-1} \sim q^2+1/L^2$ and cut off $q_y$ integrals at $q_y \sim
\pi/L$. 
Because we are interested in length scales long compared to mean free paths
we write
\begin{equation}
E_y (x,y) = \rho_H \frac{I}{L} B_{i nt}(x,y)
\end{equation}
so for the fluctuations of 
$V_H= \int dy E_y$, averaged over the contact region
$L_x$ and transformed into the frequency domain we have
\begin{equation}
\frac{\langle V_H(\omega ) \rangle^2}{\rho_H^2 I^2} =
\frac{2 f^2 \Lambda T }{ D} \Psi(\Omega)
\end{equation}
Here $\Omega = \omega L^2/(4D)$ and $\Psi$ is a dimensionless function with
the following limits: 
\begin{equation}
\Psi = \left\{ \begin{array}{ll}
\frac{1} {4 \pi^2} ln[1/\Omega] \; \; & \Omega \ll 1 \\
1/\Omega^2 & \Omega \gg 1 
\end{array} \right.
\label{Psi}
\end{equation}
Finally, for $\omega < \gamma$ one should replace $\omega$ by $\gamma$ in
(\ref{Psi}). 
These estimates apply to one $CuO_2$ layer.
Real systems have many layers and the gauge field fluctuations are
uncorrelated from layer to layer.  The voltages in the different layers add
incoherently, as may most easily be seen
by considering an experimental
configuration in which the transverse voltage difference is 
held fixed at zero
and the current fluctuation is measured.
Thus in a system of $N$ layers $\langle V^2 \rangle$ is reduced by a factor of
$1/N$.
 
We now estimate the size of a typical value of the instantaneous voltage
difference, $V_{inst}$ in a physically realizable samples.
We express this in terms of the uniform applied field, 
$B_{inst}$ which would
lead to the same voltage.
Assuming that all relevant frequencies are less
than $T$ we find
\begin{eqnarray}
B_{inst}^2 &=& \frac{V_{inst}^2}{(\rho_H I)^2} = \int \frac{d\omega}{2 \pi}
\frac{\langle V_\omega^2 \rangle}{(\rho_H I)^2} 
\nonumber \\ &=& 
\frac{f^2 \pi \Lambda T}{N L_x^2} Y(\frac{2L_x}{L_y})
\label{B_inst}
\end{eqnarray}

The final equality follows because integrating Eq. (\ref{V_H(q)}) over
frequency  produces a field correlator with short-range spatial correlations
which is moreover independent of $\gamma$.
Using the estimates previously given and setting $L_x=L_y=1 \;\mu m$, $N=200$
and $T=100K$ we find $B_{inst} \sim 0.03 \;T$.
The sample-size and cutoff $\gamma$ of course determine the frequency range in
which the noise 
power is concentrated; this range is set by the largest of $D/L^2$ (which,
using $L\sim 1 \;\mu m$ and $D\sim 10^{-2}\; cm^2/sec$, is $1\; MHz$) and
$\gamma$ which is unfortunately difficult to determine reliably because it is
an exponential function of parameters ($p_0$ and $\chi_{dia}$) which are not
well known.
The situation is further complicated because in realistic models $p_0$
varies significantly along the spinon Fermi surface
The range of plausible values of these parameters leads to estimates of
$\gamma$ ranging from $100\; kHz$ to $1 \; GHz$.
The effects due to spin fluctuations are four orders of magnitude smaller
because of the difference in $f$ and, also more difficult to observe because
$\gamma$, due physically to spin orbit scattering, is at least $1 \; THz$.

Similar considerations apply to the half-filled Landau level.
Because of the Coulomb interaction the $\langle \phi \phi \rangle$
correlator is \cite{Kalmeyer92,Halperin93,Altshuler94}
\begin{equation}
D(q,\omega) = \frac{\omega \sigma k^2}{\omega^2 + 
	\frac{Uk^2}{\sigma(|k|+\kappa)}}
\end{equation}

Here $\sigma=p_Fl/(4\pi)$ is the conductivity, $U=e^2/(8\pi\epsilon)$ and
$\kappa^{-1}$ is the screening length of the Coulomb interaction.
For typical samples, $\kappa^{-1}$ is comparable to the device size, but it
may be possible to introduce screening, e.g. via a metallic gate, so 
$\kappa^{-1} \sim l_0$. 
Our previous arguments then imply that 
\begin{equation}
B_{typ} \sim f \left( \frac{Tl_0}{U} \right)^{1/2} \left(\frac{l_0}{L}\right)
(l_0 \kappa)^{1/2}
\label{B_typ}
\end{equation}
Using $\epsilon \sim 13$ and $l_0=300\;\AA$ yields $U/l_0 \sim 2K$, so for a 
device of size $1 \; \mu m$  the effective field 
$B_{typ} [gauss] \sim  200 (T[K])^{1/2} (l_0 \kappa)^{1/2}$.
Thus for unscreened samples $\kappa^{-1} \sim 1 \; \mu m$ so at $T\sim 1\;K$, a
typical field is less than $100 \;gauss$ while for the screened samples 
$B_{typ} \sim 200 \;gauss$.

There are two contributions to the $\nu=1/2$ Hall noise: usual particle number
fluctuations and statistical gauge field fluctuations at fixed particle
number; up to a numerical factor they give identical contribution to Eq. 
(\ref{B_typ}).
The statistical flux contributions can be isolated by working at
fixed particle number (or, equivalently, by studying fluctuations of 
tan($\theta_H) =\sigma_{xy}/\sigma_{xx}$) or from the frequency dependence.
Both contributions involve the diffusion timescale which 
is $\nu_L=U/(\kappa L^2)$; the statistical contribution involves instanton
time scale $\gamma$ which dominates for large samples.
Observation of fluctuations on a time scale faster than the diffusion time
scale would be clear evidence for statistical gauge field effects.
For an unscreened sample $\kappa L \sim 1$ so for a $1 \; \mu m$ sample 
 $\nu_L \sim 150\; MHz$, while for a sample with 
$\kappa l_0 \sim 1$ it would be $\nu_L \sim 5\; MHz$.
Although instanton effects have been addressed \cite{Kim94} we believe
that the consequences of the absence of an underlying lattice have not been
adequately explored so there is no reliable estimate of $\gamma$.

To summarize: fluctuations of internal gauge fields lead to
fluctuations in the zero-applied-field Hall voltage
of high $T_c$ superconductors and in the Hall voltage of half-filled
Landau levels. 
Measurement of these fluctuations would be an important test of the validity
of the gauge theories proposed for these systems.
Their frequency dependence contains information on the
dynamics of the gauge field and in particular on instanton effects.

We thank Boris Altshuler for many useful discussions and for his 
participation in the initial stages of this work; GL acknowledges the support
by RFFI grant 950205883a and thanks  NEC Research Institute for 
hospitality.

\end{multicols}
\end{document}